\documentstyle[preprint,aps]{revtex}

\newcommand{\be}{\begin{equation}}
\newcommand{\ee}{\end{equation}}
\renewcommand\vec[1]{\mbox{\boldmath $#1$}}

\begin{document}
\title{Interface optical phonons in spheroidal dots: Raman selection rules}
\author{F. Comas$^{\ast }$, C. Trallero-Giner\thanks{%
Permanent Address: Departamento de F\'{\i}sica Te\'{o}rica, Universidad de
la Habana, Vedado 10400, Havana, Cuba.}, Nelson Studart and G. E. Marques}
\address{Departamento de Fisica, Universidade Federal de S\~{a}o Carlos,\\
13565-905, S\~{a}o Carlos, SP, Brazil.}
\maketitle

\begin{abstract}
The contribution of interface phonons to the first order Raman scattering in
nanocrystals with non spherical geometry is analyzed. Interface optical
phonons in the spheroidal geometry are discussed and the corresponding Fr%
\"{o}hlich-like electron-phonon interaction is reported in the framework of
the dielectric continuum approach. It is shown that the interface phonon
modes are strongly dependent on the nanocrystal geometry, particularly on
the ellipsoid's semi-axis ratio. The new Raman selection rules have revealed%
{\em \ } that solely interface phonon modes with even angular momentum are
allowed to contribute to the first order phonon-assisted scattering of
light. On this basis we are able to give an explanation for the observed low
frequency `shoulders'$\,$\ present in the Raman cross-section of several
II-VI semiconductor nanostructures.
\end{abstract}


By using microluminescence and micro-Raman measurements, nanocrystallites or
quantum dots (QD's) may be studied at an almost individual level.\cite%
{b1,b2,b3,b4} QD's\ built from different semiconductor materials (CdSe,
CdTe, PbS, CdS, etc.), imbedded in a glass matrix, were intensively
discussed in the later years and the spherical geometry was extensively
applied, particularly for the consideration of polar-optical phonons. \cite%
{b5,b6,b7,b8,b9} In most of the quoted papers an investigation of the
electron-phonon coupling was also made. The theoretical results have been
compared with experimental findings, in particular, resonant Raman
measurements were considered and the corresponding spectra for the first
order process show some structures on the low frequency side of the
principal peak (see for instance, Fig.~8 of Ref. \cite{b5}). The same kind
of structure in the Raman lineshape was also analyzed in \cite{b9}, where a
more realistic approach to the polar optical phonons was applied.\cite{b7,b8}
As it is shown in Fig. 1 these shoulders move to lower energies as the
quantum dot radius decreases. It{\em \ }has been claimed that the `shoulder'$%
\,$ on the left of the main Raman peak is due to surface optical (SO)-phonon
assisted transitions. However, it can be proved that, for a purely spherical
geometry, such transitions are forbidden by selection rules.\cite{b7,b10} In
order to explain the appearance of the SO-modes, the relaxation of the
angular momentum $l=0$ phonon selection is invoked. The SO-mode can be
observed in the Raman scattering processes due to: a) impurity\ or interface
imperfections, b) valence band mixing, and c) non-spherical geometry of the
QDs. For this reason, the motivation of the present communication is to
study the QD's geometrical shape deviation from the strictly spherical
geometry and its contribution to the Raman measurements. In recent works,
the electronic energy levels and wave functions of spheroidal QD's were
examined.\cite{b11} In the current report we consider the polar-optical
vibrations of a QD with spheroidal geometry by applying the dielectric
continuum approach. We{\em \ }study the changes introduced in the SO-phonon
eigenfrequencies, eigenstates and also in the electron-phonon Hamiltonian
due to non spherical geometry. The key point is to analyzing the selection
rules for the first order phonon-assisted Raman scattering as a function of
the QD geometry. The conclusion that solely SO-phonons with even angular
momentum are allowed to have a contribution to the Raman spectra provides us
with a strong{\em \ }basis for the explanation of the origin{\em \ }of the
`shoulders'$\,$ already invoked in previous works.

Let us briefly summarize the essential theory that{\em \ }we have applied.
The main macroscopic quantities involved in the description of polar-optical
vibrations, in particular the involved electric potential $\varphi $, is
derived from the equation $\epsilon (\omega )\nabla ^{2}\varphi =0$, where
the frequency dependent dielectric function is given by $\epsilon (\omega
)=\epsilon _{\infty }(\omega ^{2}-\omega _{L}^{2})/\omega ^{2}-\omega
_{T}^{2})$. For the SO-phonons $\epsilon (\omega )\neq 0$ thus, the solution
of Laplace equation $\varphi $ : {\em \ }i) should be continuous at the
interface between two different media, and{\em \ }ii) must fulfil the
boundary condition $\epsilon _{1}[\partial \varphi _{1}/\partial
n]_{S}=\epsilon _{2}[\partial \varphi _{2}/\partial n]_{S}$. By considering
a prolate spheroidal QD, the coordinates $\xi \,,\,\eta \,,\,\phi $ are the
most convenient, and related to the rectangular Cartesian coordinates
through the equations\cite{b12,b13}: $x=b\sqrt{(\xi ^{2}-1)(1-\eta ^{2})}%
\cos \phi $ , $y=b\sqrt{(\xi ^{2}-1)(1-\eta ^{2})}\sin \phi $ and $z=b\xi
\eta $. We also have $\xi \geq 1$, $-1\leq \eta \leq 1$ and $0\leq \phi \leq
2\pi $. The equation $\xi =$ cons. describes an ellipsoid of revolution
where the $z$ direction is taken along the ellipsoid's major axis with{\em \ 
}$2b${\em \ }being{\em \ }the interfocal distance. For $1\leq \xi \leq \xi
_{0}$ we have, in the ellipsoid's interior region, a semiconductor of the
CdSe prototype with a dielectric function $\epsilon (\omega )$. For $\xi
\geq \xi _{0}$ we shall consider a glass matrix with frequency independent%
{\em \ }dielectric constant $\epsilon _{D}$. The Laplace equation is
separable in the spheroidal prolate coordinates and the solutions are given
by\cite{b12} 
\begin{eqnarray}
\varphi ^{<} &=&A_{lm}R_{l}^{m}(\xi )Y_{lm}(\eta \,,\,\phi ),\ \ \text{for}%
\,\ \ \xi \leq \xi _{0},  \nonumber  \label{e1} \\
\varphi ^{>} &=&A_{lm}\left( R_{l}^{m}(\xi _{0})/Q_{l}^{m}(\xi _{0})\right)
Q_{l}^{m}(\xi )Y_{lm}(\eta \,,\,\phi ),\quad \text{for}\quad \xi \geq \xi
_{0}.
\end{eqnarray}%
Notice that the potential is already continuous at $\xi =\xi _{0}$. The
other boundary condition is fulfilled by taking $\epsilon _{1}\equiv
\epsilon (\omega )$ and $\epsilon _{2}\equiv \epsilon _{D}$, which leads to
the following result: 
\begin{equation}
\frac{\epsilon (\omega )}{\epsilon _{D}}=\left( \frac{d}{d\xi }\ln
Q_{l}^{m}|_{\xi _{0}}\right) \left( \frac{d}{d\xi }\ln R_{l}^{m}|_{\xi
_{0}}\right) ^{-1}\equiv f_{lm}(\xi _{0}),  \label{e2}
\end{equation}%
where the functions $R_{lm}${\em \ }and{\em \ }$Q_{lm}${\em \ }are defined
below. The functions $f_{lm}{\em (\xi }_{0}{\em )}$ are neither{\em \ }%
dependent on the nature of the constituent materials nor of the
normalization of the functions $R_{lm}$ and $Q_{lm}$. They do{\em \ }depend
on the QD geometrical shape through the parameter $\xi _{0}$. The SO-phonons
eigenfrequencies in the spheroidal QD are then given by%
\begin{equation}
\frac{\omega _{lm}^{2}}{\omega _{T}^{2}}=\frac{\epsilon _{0}-\epsilon
_{D}f_{lm}(\xi _{0})}{\epsilon _{\infty }-\epsilon _{D}f_{lm}(\xi _{0})}.
\label{e3}
\end{equation}%
It is easy to show that the limit{\em \ }$\xi _{0}\rightarrow \infty $ in
Eq.(\ref{e3}) leads to the corresponding eigenfrequencies of a purely
spherical QD.\cite{b5}

The functions $R_{l}^{m}(\xi )$ and $Q_{l}^{m}(\xi )$ are defined in Ref. %
\cite{b12} and we shall give them here in terms of hypergeometric functions%
\begin{eqnarray}
R_{l}^{m}(\xi ) &=&\frac{(2l)!(\xi ^{2}-1)^{m/2}\xi ^{l-m}}{2^{l}l!(l-m)!}F%
\left[ \frac{m-l}{2}\,,\,\frac{m-l+1}{2}\,,\,\frac{1}{2}-l\,,\,\frac{1}{\xi
^{2}}\right] ,  \nonumber  \label{e4} \\
Q_{l}^{m}(\xi ) &=&\frac{2^{m}(l-m)!\Gamma (1/2)(\xi ^{2}-1)^{m/2}}{\Gamma
(l+3/2)(2\xi )^{l+m+1}}F\left[ \frac{l+m+1}{2}\,,\,\frac{l+m+2}{2}\,,\,\l +%
\frac{3}{2}\,,\,\frac{1}{\xi ^{2}}\right] \,.
\end{eqnarray}%
The functions $Y_{lm}(\eta ,\phi )$ are the usual spherical harmonics and,
in all cases, the quantum numbers are given by $l=1,2,3,\cdots $ and $%
|m|\leq l$. The other limit properties are: i){\em \ }$R_{l}^{m}(\xi )$ is
divergent as $\xi ^{l}$ when $\xi \rightarrow \infty $ and{\em \ }are
convergent at $\xi =1$; ii) $Q_{l}^{m}(\xi )$ converges to zero as $\xi
^{-l-1}$ when $\xi \rightarrow \infty $ and diverges logarithmically at $\xi
=1$. We assume $\xi _{0}>1$.

The corresponding electron-phonon Hamiltonian $\hat{H}_{e-ph}(\vec{r})=-e%
\hat{\varphi}(\vec{r})$ can be derived by standard procedures. The potential
operator $\hat{\varphi}$ can be written as%
\begin{equation}
\hat{\varphi}(\vec{r})=\sum_{lm}\frac{\epsilon _{\infty }-\omega _{L}}{%
\epsilon _{\infty }-\epsilon _{D}f_{lm}(\xi _{0})}\left[ \frac{2\pi \hbar }{%
\epsilon ^{\ast }b\omega _{lm}g_{lm}(\xi _{0})}\right] ^{1/2}\left\{
F_{l}^{m}(\xi )Y_{lm}(\eta \,,\phi )\hat{a}_{lm}+\text{h.c}.\right\} ,
\label{e5}
\end{equation}%
where $F_{l}^{m}=R_{l}^{m}(\xi )$ for $\xi \leq \xi _{0}$ and $F_{l}^{m}=%
\left[ R_{l}^{m}(\xi _{0})/Q_{l}^{m}(\xi _{0})\right] Q_{l}^{m}(\xi )$ for $%
\xi \geq \xi _{0}$. Moreover, $1/\epsilon ^{\ast }=(1/\epsilon
_{0}-1/\epsilon _{\infty })$.

Let us consider the case of a CdSe spheroidal QD imbedded in a glass matrix.
The applied physical parameters are $\omega _{T}=165.2\,$cm$^{-1}$, $%
\epsilon _{0}=9.53$, $\epsilon _{\infty }=5.72$, while, for the host
material,{\em \ }we take $\epsilon _{D}=4.64$.\cite{b9} For prolate
ellipsoidal geometry{\em \ }the phonon frequencies $\omega _{lm}$ as a
function of the deviation parameter $\xi _{0}$ are presented in Fig.~2a, for 
$l=1,\,2$. Each involved SO-phonons modes are explicitly indicated in the
figure. The dotted lines are the corresponding eigenfrequencies for the
strictly spherical case. In Fig.~2b we are showing the same dependencies for 
$l=3$. Notice the splitting of the frequencies (according to the rule $m\leq
l$){\em \ }and the main conclusion is that the separation between {\it %
SO-phonon} {\it frequencies depends on the QD dimensions} (through $\xi _{0}$%
). We have found that the observed frequency splitting is in the range of
the structural features seen in the spectral lineshapes of Fig. 1. For
higher values of $l$ we obtain lower values for the frequency splitting.
Another{\em \ }important quantity is the ellipsoid's semi-axes ratio $r=\xi
_{0}/\sqrt{\xi _{0}^{2}-1}$. According to Ref. \cite{b14} we should expect a
ratio $1.1\leq r\leq 1.3$. A direct comparison between the experimental data
and the results here presented should provide a much{\em \ }better
understanding of the{\em \ }role played by the{\em \ }QD geometry.

First-order resonant Raman scattering cross-sections of a single QD is
proportional to the square of the scattering amplitude{\em ,} $W_{FI}${\em ,}
between the initial and final states, $I$ and $F$, as{\em \ }given by

\begin{equation}
W_{FI}=\sum\limits_{\mu _{1},\mu _{2}}\frac{\left\langle F\left|
H_{E-R}^{+}\right| \mu _{2}\right\rangle \left\langle \mu _{2}\left|
H_{E-P}\right| \mu _{1}\right\rangle \left\langle \mu _{1}\left|
H_{E-R}^{-}\right| I\right\rangle }{(\hbar \omega _{s}-E\mu _{2})(\hbar
\omega _{l}-E\mu _{1})}.  \label{prob}
\end{equation}%
Here, $\omega _{l}$ ($\omega _{s})$ is the incoming (scattered) and $H_{E-P}$
($H_{E-R}$) is the electron-hole phonon (electron-radiation) Hamiltonian
interaction. The corresponding electron-hole wave functions, $\left| \mu
\right\rangle ,$ were taken in the same spirit of Ref. \cite{b11}, but
extended to the QD exterior region, i.e., hard wall boundary conditions on
the spheroid's surface were not assumed. By introducing Eq. (\ref{e5}) into
Eq. (\ref{prob}) we were able to obtain selection rules for the
electron-phonon transitions, which are summarized as follows: (a) Only
SO-phonons with $m=0$ and $l=\;$even integer are allowed. Notice that $l=1$
is not allowed for the transitions in contradiction to the assumptions of
previous works;\cite{b9} (b) For the electronic states (denoted as in Ref. %
\cite{b11}) the angular momenta{\em \ }$l_{e}$ and $l_{h}$ should have the
same parity, while $m_{e}=m_{h}=m$. By $e$ ($h$) we mean electron (hole)
quantum numbers. The later results permit us to give an interpretation for
the `shoulder'$\,$ at the left side of the main Raman peak seen in Fig. 1 as
a direct consequence of the spheroidal geometry of the dot. On the other
hand, the observed Raman data, together with the spheroidal SO-phonons here
reported, can be used in order to determine the ellipsoid's semi-axes ratio $%
r$. On the basis of data taken from Ref. \cite{b9}, we have that, for the QD
of Fig. 1(a) where the `shoulder' maximum is seen at{\em \ }approximately
183 cm$^{-1}$, the corresponding SO-phonon frequency for the prolate QD
(setting $l=2$ and $m=0$) gives a ratio{\em \ } $r=1.86$. This result
indicates that this QD has a very strong deviation from the spherical
geometry. On the other hand, the `shoulder' maximum at 188 cm$^{-1}$ in Fig.
1(b) with mean radius of 2.6 nm will lead to $r=1.065$, an indication that
this is a sample with a shape closer to spherical geometry. The later
results confirm the general idea that QD's with large mean radius\ should
display a more spherical shape.

We acknowledge financial support from Funda\c{c}\~{a}o de Amparo \`{a}
Pesquisa do Estado de S\~{a}o Paulo(FAPESP) and Conselho Nacional de
Desenvolvimento Cient\'{\i}fico e Tecnol\'{o}gico (CNPq). F. C. and C. T. G.
are grateful to Departamento de F\'{\i}sica, Universidade Federal de S\~{a}o
Carlos, for\ the hospitality.

\begin{figure}[tbp]
\caption{First-order Raman lineshape for CdSe nanocrystal from Ref. %
\protect\cite{b9}. (a) Mean radius $R_{0}=1.8$ nm. (b) Mean radius $R_{0}=2.6
$ nm. The solid lines correspond to the calculation of the Raman spectrum
assuming a QD with spherical geometry. Dots represent the spectra
measurement.}
\label{1}
\end{figure}

\begin{figure}[tbp]
\caption{(a) Squared frequencies $\protect\omega _{lm}^{2}$ in units of $%
\protect\omega _{T}^{2}$ as a function of $1/\protect\xi _{0}$ for $l=1,\,2$
and all possible values of $m$ for the prolate ellipsoid. (b) Same plot for $%
l=3$ and all possible values of $m$. In both cases the dotted lines
correspond to the strictly spherical case. }
\label{2}
\end{figure}


\begin{references}
\bibitem{b1} A. Ekimov, J. Lumin. {\bf 70}, 1 (1996).

\bibitem{b2} D. Bertram, Phys. Rev. B {\bf 57}, 4265 (1998).

\bibitem{b3} S. A. Empedocles, D. J. Norris, and M. G. Bawendi, Phys. Rev.
Lett. {\bf 77}, 3877 (1996).

\bibitem{b4} D. J. Norris, A. Sacra, C. B. Murray, and M. G. Bawendi, Phys.
Rev. Lett. {\bf 72}, 2612 (1994).

\bibitem{b5} M. C. Klein, H. Hache, D. Ricard and C. Flytzanis, Phys. Rev. B 
{\bf 42}, 11123 (1990).

\bibitem{b6} S. Nomura and T. Kobayashi, Phys. Rev. B {\bf 45}, 1305 (1992).

\bibitem{b7} E. Roca, C. Trallero-Giner, and M. Cardona, Phys.Rev.B {\bf 49}%
, 13704 (1994).

\bibitem{b8} M. P. Chamberlain, C. Trallero-Giner, and M. Cardona, Phys.
Rev. B {\bf 51}, 1680 (1995).

\bibitem{b9} C. Trallero-Giner, A. Debernardi, M. Cardona, E.
Menendez-Proupin, and A. I. Ekimov, Phys. Rev. B {\bf 57}, 4664 (1998).

\bibitem{b10} E. Duval, Phys.Rev B {\bf 46}, 5795 (1992).

\bibitem{b11} G. Cantele, D. Ninno, and G. Iadonisi, J. Phys.:
Condens.Matter {\bf 12}, 9019 (2000) and references therein.

\bibitem{b12} P. M. Morse and H. Feshbach, {\it Methods of Theoretical
Physics} (Mc.Graw-Hill, New York, 1953).

\bibitem{b13} {\it Handbook of Mathematical Functions, }edited by{\it \ }M.
Abramowitz and I. Stegun, (Dover, New York, 1972) p. 751.

\bibitem{b14} N. Nirmal, D. J. Norris, M. Kuno, M. G. Bawendi, Al. L. Efros,
and M. Rosen , Phys. Rev. Lett. {\bf 75}, 3728 (1995).
\end{references}
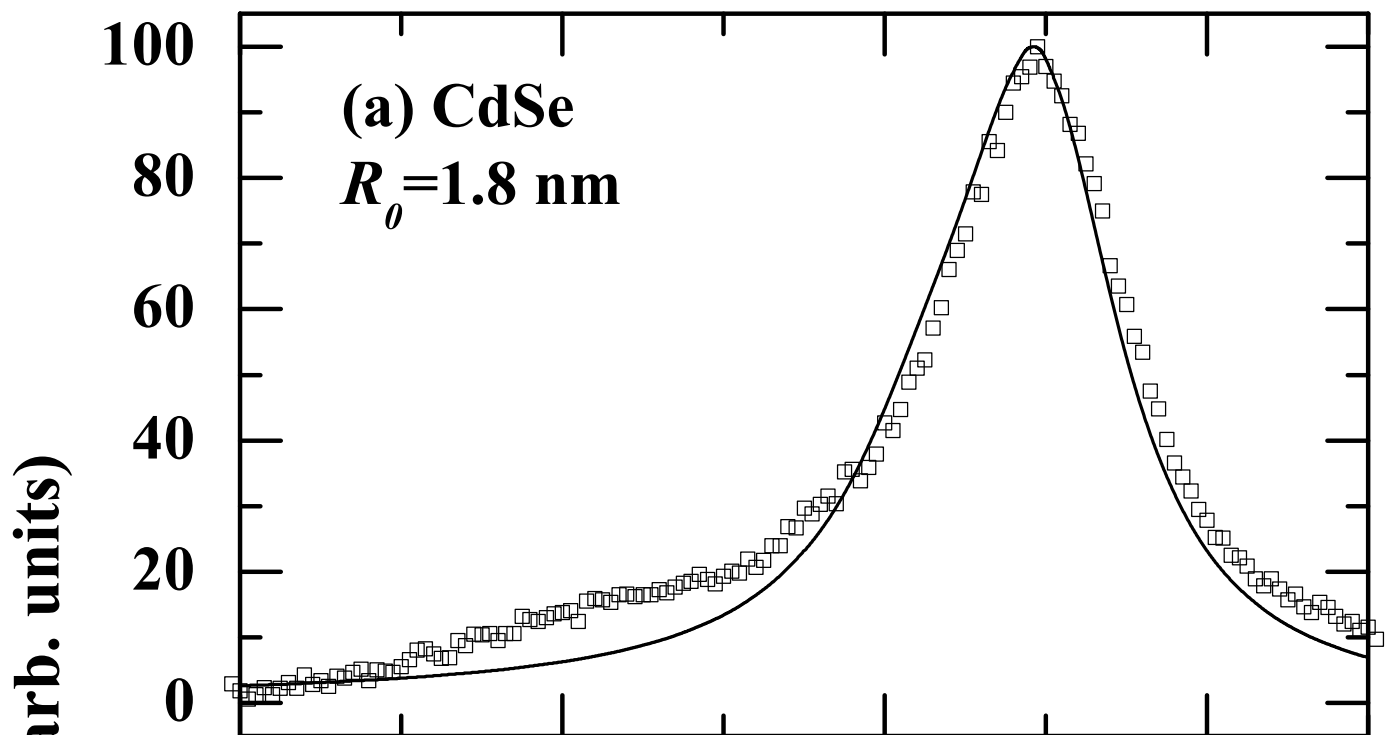
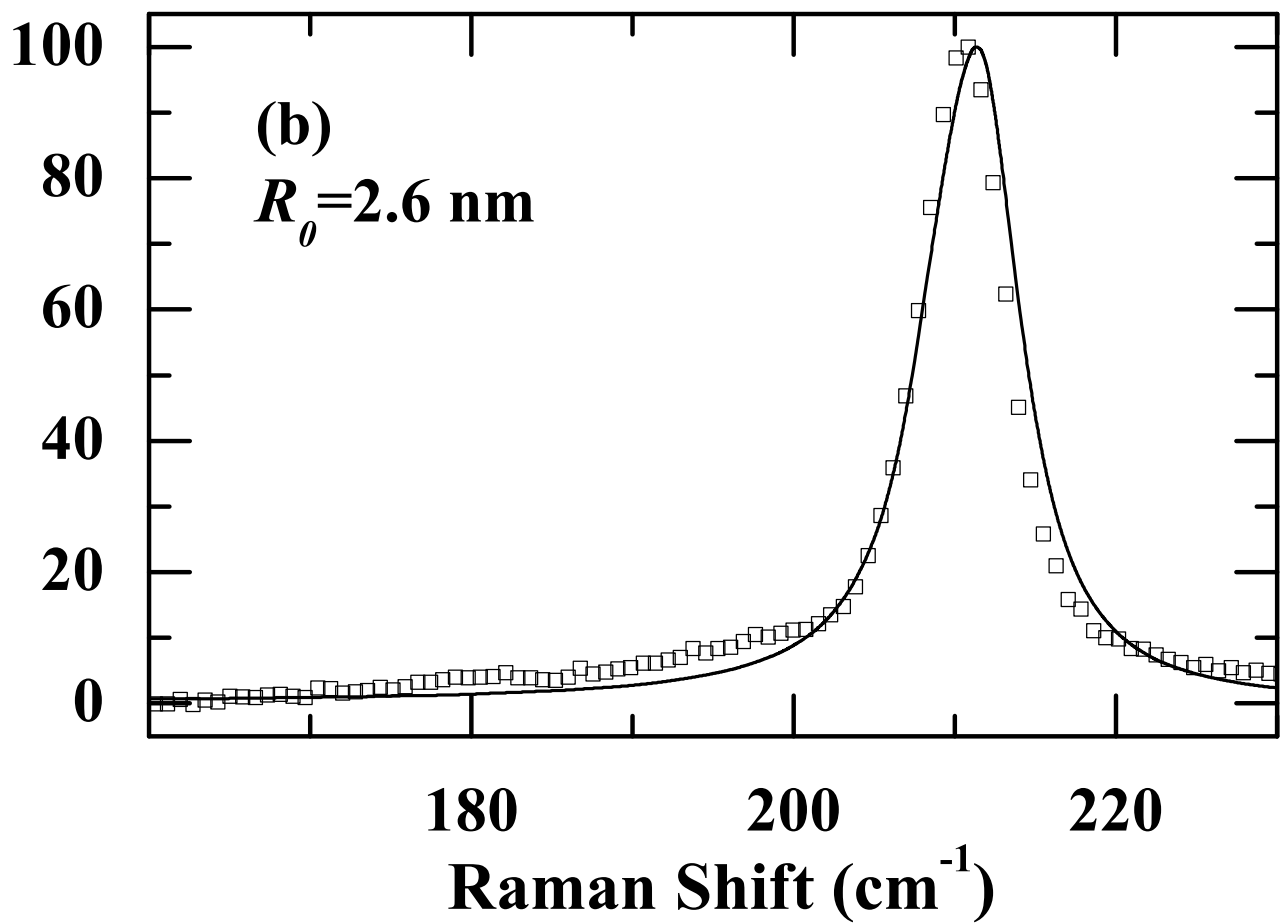
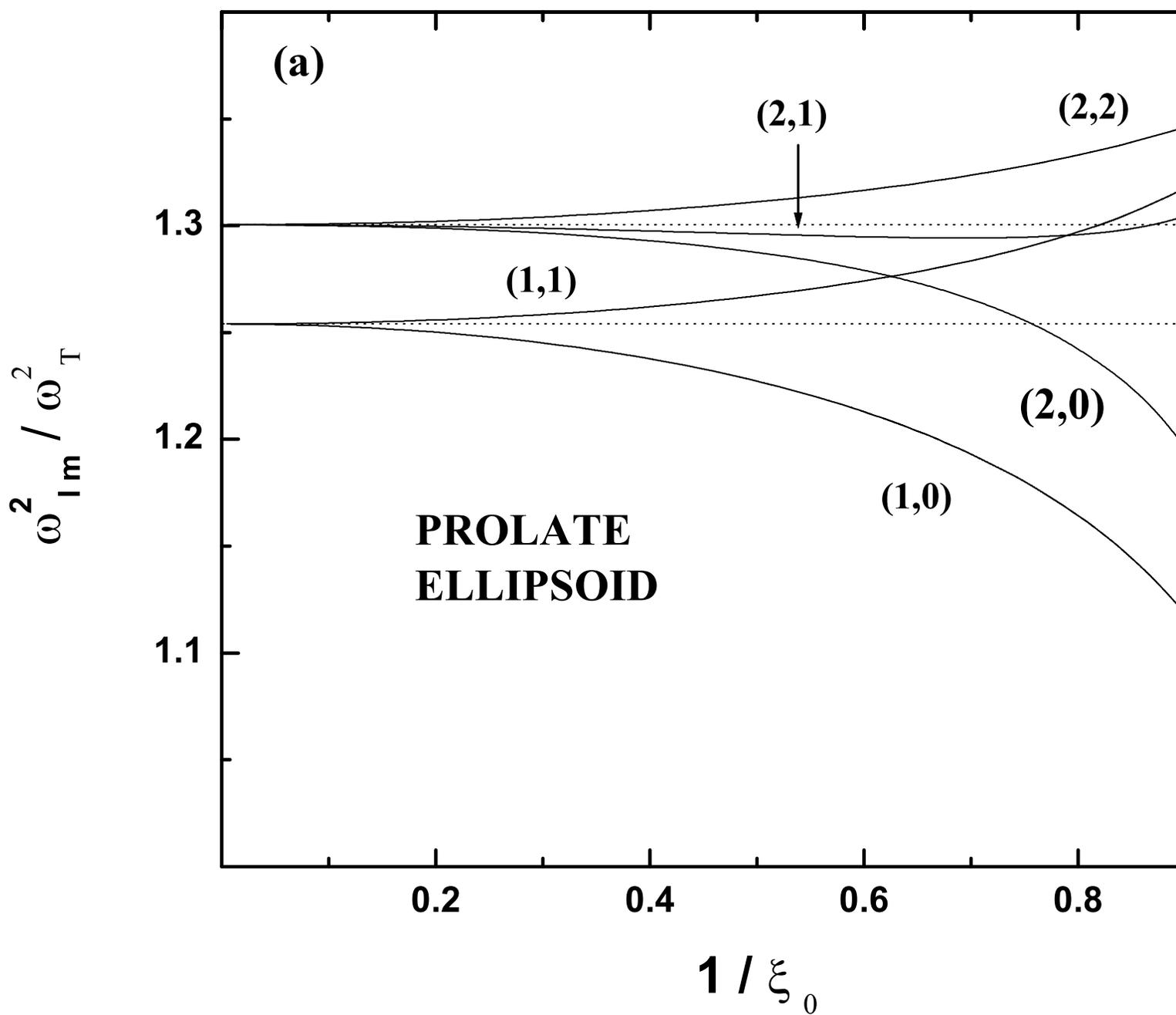
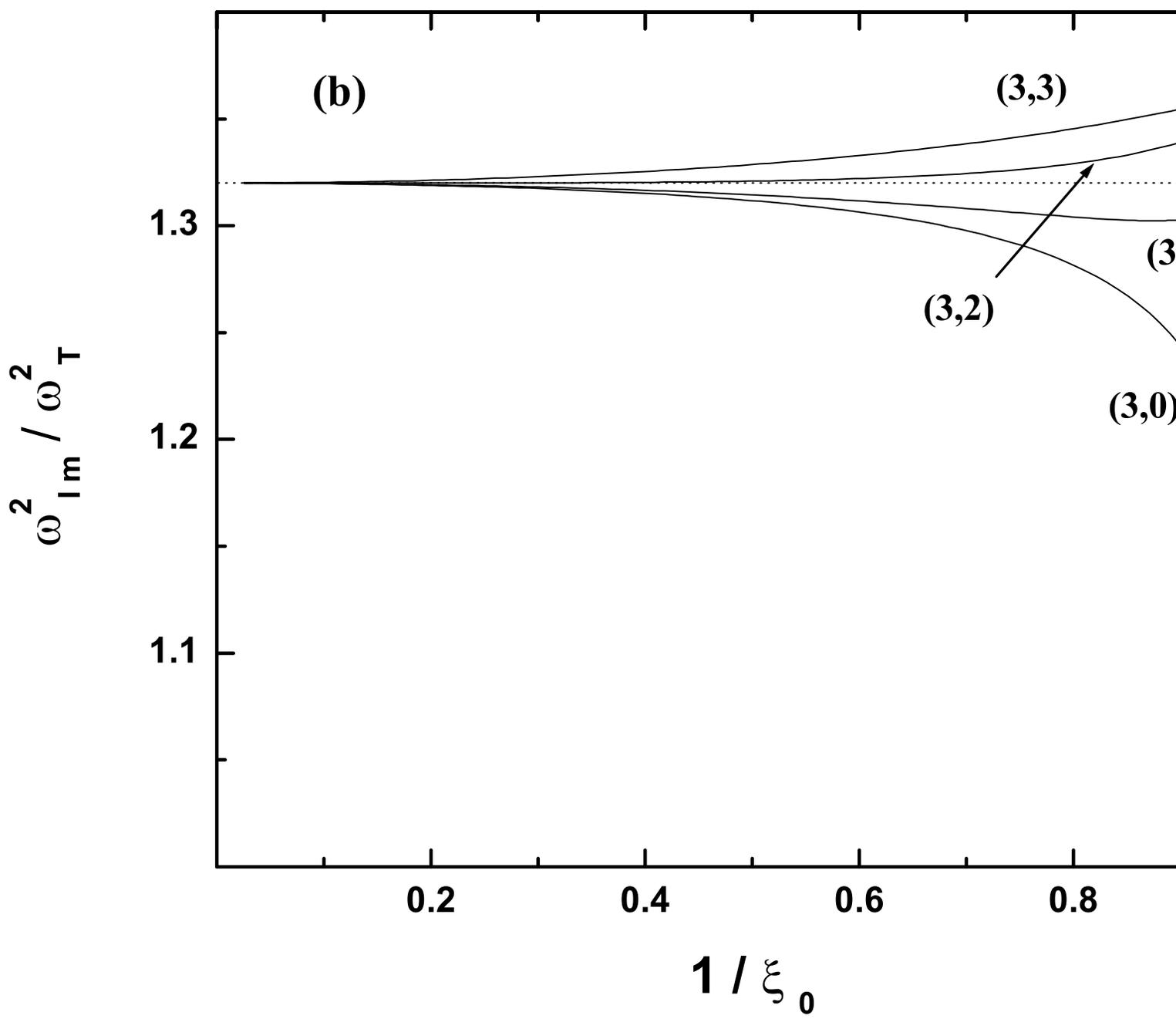
\end{document}